# In Vivo Magnetic Resonance Spectroscopy by J-Locked Chemical Shift Encoding for Determination of Neurochemical Concentration and Transverse Relaxation Time

Li An and Jun Shen

National Institute of Mental Health, National Institutes of Health, Bethesda, MD, USA

Corresponding author:

Li An

Building 10, Room 3D46

10 Center Drive, MSC 1216

Bethesda, MD 20892-1216

Voice: (301) 594-6868

Email: li.an@nih.gov

# Abstract

Cell pathology in neuropsychiatric disorders has mainly been accessible by analyzing postmortem tissue samples. Although molecular transverse relaxation informs local cellular microenvironment via molecule-environment interactions, precise determination of the transverse relaxation times of molecules with scalar couplings (J), such as glutamate and glutamine, is difficult using current in vivo magnetic resonance spectroscopy (MRS) technologies, whose approach to measuring transverse relaxation has not changed for decades. We introduce an in vivo MRS technique that achieves chemical shift encoding with selectively locked J-couplings in each column of the acquired two-dimensional dataset, freeing up the entire row dimension for transverse relaxation encoding. This results in increased spectral resolution, minimized background signals, and markedly broadened dynamic range for transverse relaxation encoding. This technique enables determination of the transverse relaxation times of glutamate and glutamine in vivo with unprecedented high precision. Since glutamate predominantly resides in glutamatergic neurons and glutamine in glia in the brain, this noninvasive technique provides a way to probe cellular pathophysiology in neuropsychiatric disorders for characterizing disease progression and monitoring treatment response in a cell type-specific manner in vivo.

## Introduction

Magnetic resonance spectroscopy (MRS) enables noninvasive detection of molecules in vivo, including those found in the highly inaccessible human brain. Neuropsychiatric disorders and brain tumors are among the most active areas of in vivo MRS research. Accurate in vivo detection of key neurochemicals such as glutamate (Glu), glutamine (Gln), and glutathione (GSH) is highly important in these fields. Glu is the primary excitatory neurotransmitter in the central nervous system (CNS). The metabolic coupling between Glu and Gln forms the Glu-Gln neurotransmitter cycle between glutamatergic neurons and glia[1, 2]. Glu and Gln are also major metabolites in the CNS, playing crucial roles in normal brain function and cancer cell growth[3, 4]. GSH is an antioxidant, whose level is a marker of redox state[5]. Glu, Gln, and GSH have been implicated in many illnesses such as epilepsy, schizophrenia, bipolar disorders, Alzheimer's disease, major depressive disorder, anxiety disorders, and brain tumors[6-11]. Many brain disorders are also associated with alterations in cell type-specific microenvironments[12], which are difficult to measure noninvasively. For instance, postmortem studies have found ample evidence of glial pathology in major depressive disorder[13, 14], and one of the defining characteristics of the pathophysiological state of Alzheimer's disease is the atrophy of glutamatergic neurons[15].

For the majority of in vivo single-voxel MRS experiments, localized one-dimensional free induction decay (FID) signals are acquired with a large number of transients to achieve sufficient signal-to-noise ratio. Chemical shift information is frequency-encoded in the FIDs, along with signal modulation due to J-coupling. The frequency domain MRS spectrum is obtained by Fourier transforming the sum of all transients. Signal modulation due to J-coupling causes peak splitting in an MRS spectrum, which can impair quantification by reducing peak amplitudes and increasing spectral overlap among different molecules and with the background signals (macromolecule signals and the background spectral baseline). Recent MRS studies of the human brain have shown that the background signals can cause significant errors in quantifying the concentrations of J-coupled neurochemicals using the widely used short echo time (TE) MRS[16].

In addition to measuring concentrations, in vivo MRS can also determine transverse relaxation times ($T_2$). Unlike the $T_2$ of tissue water measured by magnetic resonance imaging (MRI), the molecular $T_2$s measured by MRS are often cell type-specific and can provide critical information on the cellular microenvironment where the molecules reside through molecule-local environment interactions[17-21]. For instance, in the brain, Glu and Gln are predominantly localized in glutamatergic neurons and glia, respectively. Reliable detection of Glu and Gln $T_2$s can therefore provide unique information on the microenvironments of these different cell types in vivo.

There are several inherent difficulties in reliably measuring $T_2$ of J-coupled molecules: (i) the signal intensity of many J-coupled molecules is low, partly due to peak splitting caused by J-couplings; (ii) due to J-evolution and spectral overlap, the optimal TE for reliable detection of a J-coupled molecule is fixed[22, 23], while $T_2$ measurements require a series of very different time points for transverse relaxation encoding; (iii) the background signals also change with TE[24], and the uncertainty in modeling the background signals leads to increased errors in $T_2$ measurements.

In this work, we introduce a two-dimensional (2D) MRS technique for measuring the concentrations and $T_2$ relaxation times of neurochemicals in vivo. Unlike traditional 2D nuclear magnetic resonance (NMR) spectroscopy techniques, this technique does not involve coherence or polarization transfer. The column ($t_1$) dimension is used for chemical shift encoding and the row ($t_2$) dimension is used for $T_2$ encoding. By using frequency-selective J-locking pulses, the targeted J-evolutions are locked in each column of the 2D dataset. The Fourier transform of each column of the 2D dataset produces a spectrum with the targeted peak splitting eliminated. This J-locked chemical shift encoding (JL-CSE) technique addresses the difficulties in measuring the concentrations and T$_2$s of J-coupled molecules by: (i) selectively decoupling the targeted spins in the column dimension; (ii) allowing T$_2$ encoding at many time points with markedly broadened dynamic range; (iii) utilizing the TE averaging effect[25, 26] to minimize the background signals by averaging the columns of the 2D dataset into a small number of bins. As an application of JL-CSE, the concentrations and T$_2$ relaxation times of Glu, Gln, and GSH–three important biomarkers for neuropsychiatric and brain cancer research–are measured in the human brain in vivo.

## Results

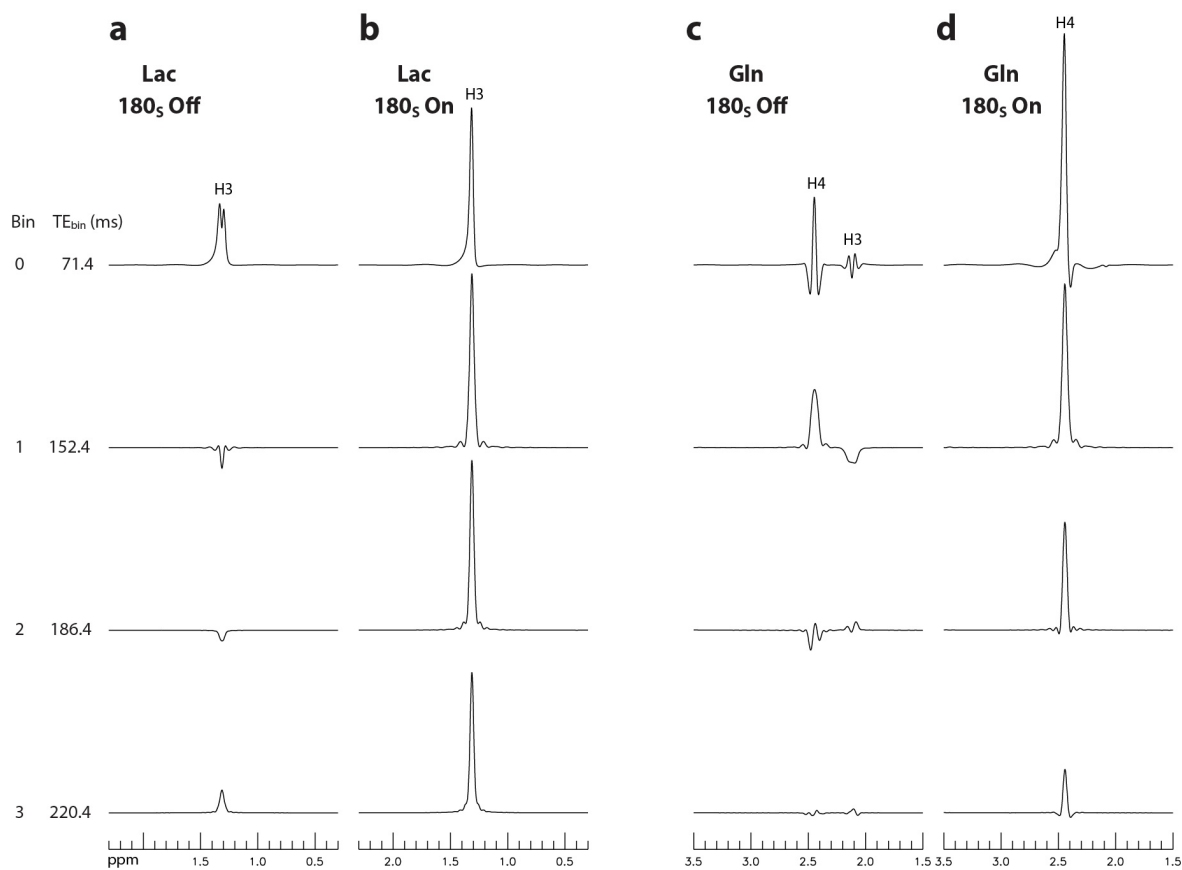

**Figure 1** Numerically calculated bin spectra of Lac (**a**, **b**) and Gln (**c**, **d**) using the pulse sequence shown in Figure 8 in the Methods section. Each stack plot displays a subset of the 11 bin spectra

obtained. The dual-band J-locking pulses (denoted as 180$_S$) had two 180° bands, one at 2.12 ppm and the other at 4.10 ppm. The J-locking pulses were turned off in **a** and **c** and on in **b** and **d**. The [Lac]:[Gln] ratio was 1:7. TE$_{bin}$ represents the average TE of each bin. Lac: lactate; Gln: glutamine.

Density matrix simulated bin spectra of lactate (Lac) and Gln are shown in Figure 1. For each molecule, two sets of bin spectra were calculated with the frequency-selective J-locking pulses switched off and on, respectively. Since there is no reported T$_2$ value for Lac in the healthy human brain, the T$_2$ of Lac was set to 180 ms, similar to that of Glu in the grey matter[23]. T$_2$ of Gln was set to 80 ms[23]. All bin spectra were line broadened to a typical in vivo linewidth of 9 Hz. Without the J-locking pulses (Figure 1a), the Lac peak is a doublet with low amplitude for bin 0 and becomes even lower for bins 1 – 3 due to its H2-H3 J-evolution. In contrast, when the J-locking pulses are turned on (Figure 1b), the frequency band at 4.10 ppm locks the H2-H3 J-evolution in the column (t$_1$) dimension. As a result, the Lac H3 peak becomes a sharp singlet in the column dimension with much higher amplitude in all four bin spectra. Gln has a more complex spin system with a strong internal coupling between its two H4 protons but weak external couplings between its H3 and H4 protons[27]. Without the J-locking pulses, the Gln H4 signal is greatly diminished due to J-evolution and signal self-cancelation (Figure 1c). By applying the J-locking pulses to the H3 resonances of Gln at 2.12 ppm, the H3-H4 J-evolutions are locked in the column dimension, resulting in a sharp pseudo singlet for the H4 signal with significantly higher peak intensity (Figure 1d). Figure 1 shows that the application of the J-locking pulses resulted in a significant improvement in both spectral resolution and peak amplitude for the targeted spins.

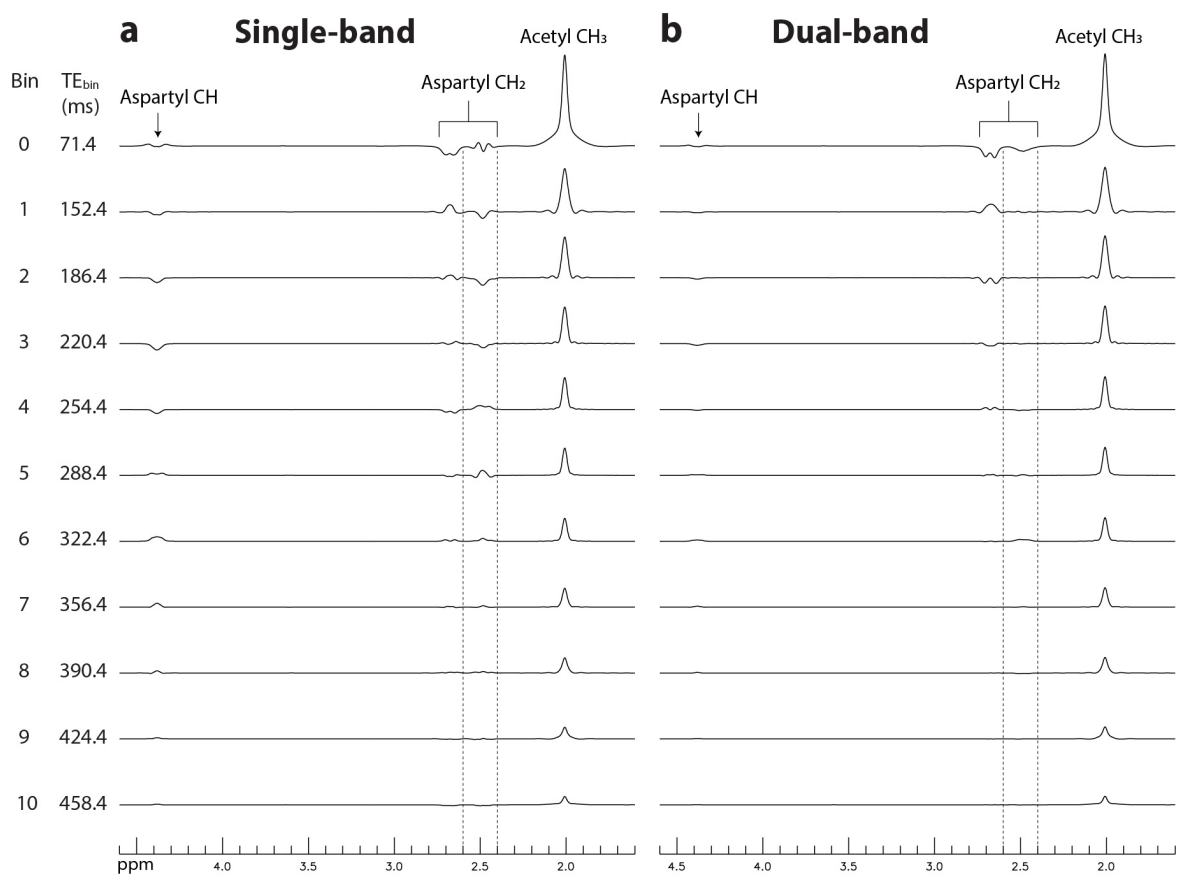

**Figure 2** Numerically calculated bin spectra of NAA obtained using single-band (**a**) and dual-band (**b**) J-locking pulses. The single-band 180° J-locking pulses were applied at 2.12 ppm. The dual-band J-locking pulses had a 180° band at 2.12 ppm and a 90° band at 4.38 ppm targeting the NAA aspartyl CH proton. The remaining parameters are provided in Figure 8 in the Methods section. The pair of dashed lines in each stack plot indicate the 2.4 – 2.6 ppm range, where Gln and glutamyl GSH H4 protons resonate. NAA: N-acetylaspartate; GSH: glutathione.

Density matrix simulated bin spectra of N-acetylaspartate (NAA) using single-band (a) and dual-band (b) J-locking pulses are shown in Figure 2. $T_2$ was set to 200 ms for acetyl $CH_3$ and 140 ms for aspartyl $CH_2$[23]. All bin spectra were line broadened to a typical in vivo linewidth of 9 Hz. In Figure 2a, the NAA aspartyl $CH_2$ resonances at 2.4 – 2.6 ppm are relatively large and interfere with the detection of Gln and GSH. In contrast, in Figure 2b, these resonances were effectively suppressed in bins 0 – 3 due to the second band targeting 4.38 ppm.

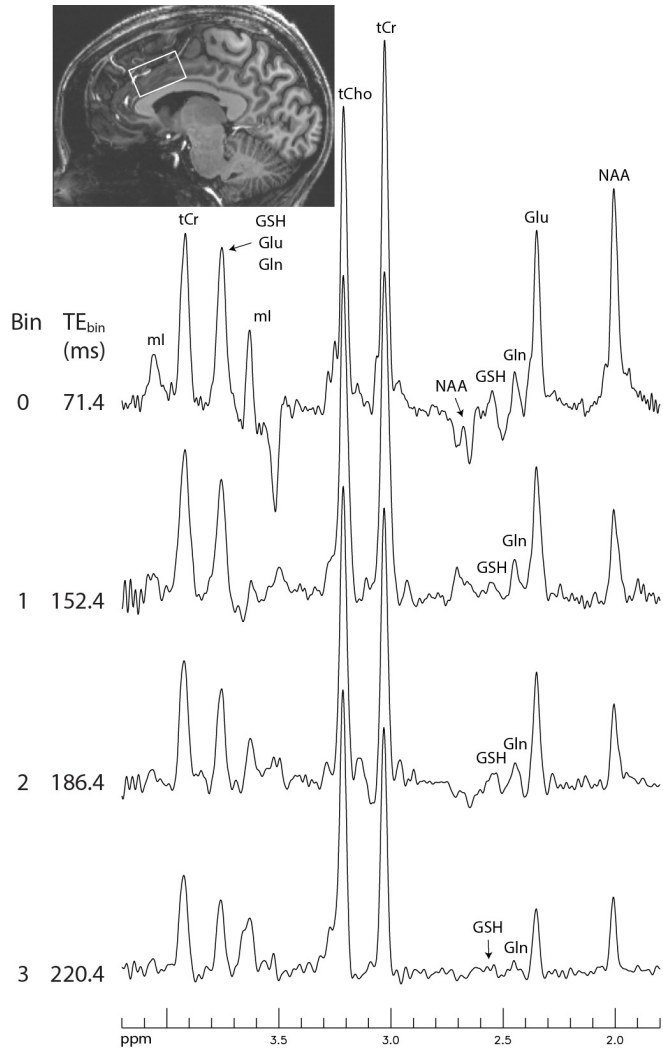

**Figure 3** In vivo bin spectra (0 - 3) acquired from a $2 \times 3.5 \times 2$ cm$^3$ voxel in the cingulate cortex of a healthy participant. The pulse sequence used to acquire the in vivo data was the same as that used for Figure 2b, employing dual-band J-locking pulses to simultaneously lock the H3-H4 couplings and suppress the NAA aspartyl resonances between 2.4 and 2.6 ppm. No line broadening was applied to the spectra. TE$_{bin}$ represents the average TE of each bin. mI: myo-Inositol; tCho:

total choline; tCr: total creatine; Glu: glutamate; Gln: glutamine; GSH: glutathione; NAA: N-acetylaspartate. The full dataset is shown in Figure 4.

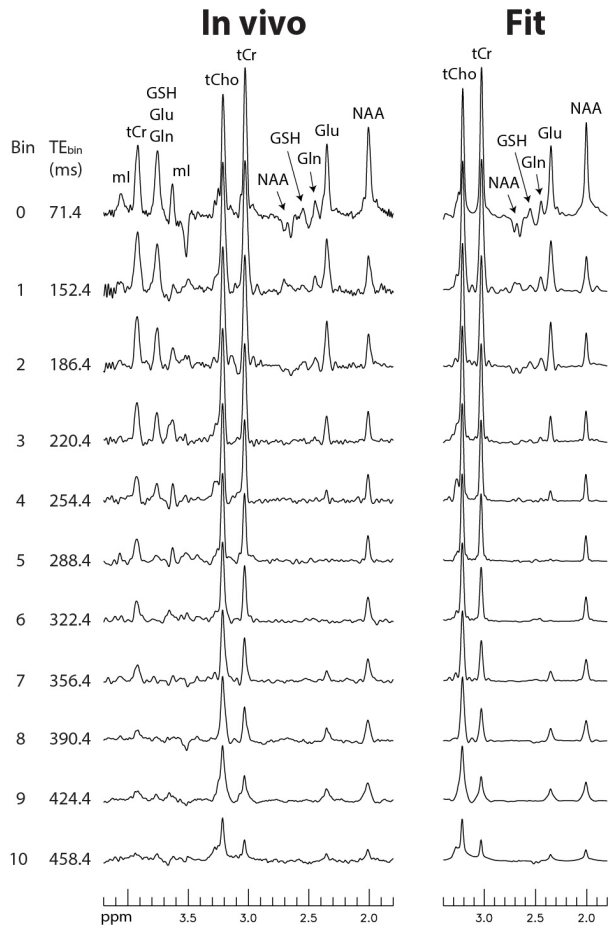

**Figure 4** In vivo bin spectra (0 - 10) and their fits. The entire scan time was 10 min and 43 s using a 2.5 s repetition time (TR).

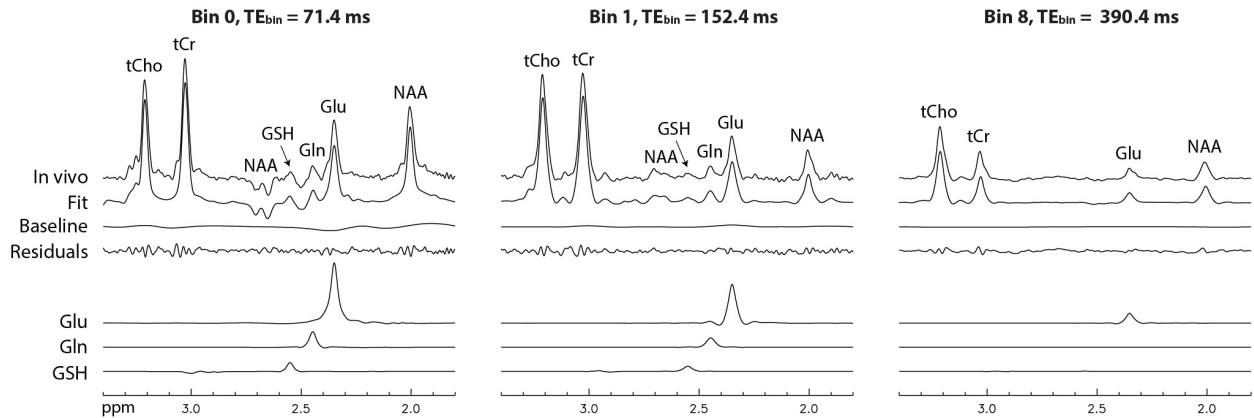

**Figure 5** Examples of fitting details for the bin spectra displayed in Figure 4.

JL-CSE was used to measure the concentrations and $T_2$s of Glu, Gln, GSH, total creatine (tCr), and total choline (tCho) in the human brain. Figure 3 shows the first four bin spectra acquired from a healthy participant and the re-test results are provided in the Supplementary Information. The Gln H4 signal appears as a sharp pseudo singlet in the bin spectra, which is consistent with the numerical simulation shown in Fig. 1d. The Gln peak remains detectable in bin 3, which has an average TE ($TE_{bin}$) of 220.4 ms. Figure 4 displays the spectra for all 11 bins and their corresponding fitted spectra. The fitted spectra in the fitting range (1.8 – 3.4 ppm) closely resemble the in vivo spectra for all 11 bins. The Glu peak appears prominent in bins 0 – 3, which span 71.4 – 220.4 ms, and is still detectable in bin 10, which has a very long $TE_{bin}$ of 458.4 ms. Figure 5 shows the fitting details for bins 0, 1, and 8. The model spectra fit the in vivo spectra well, as evidenced by the small residuals.

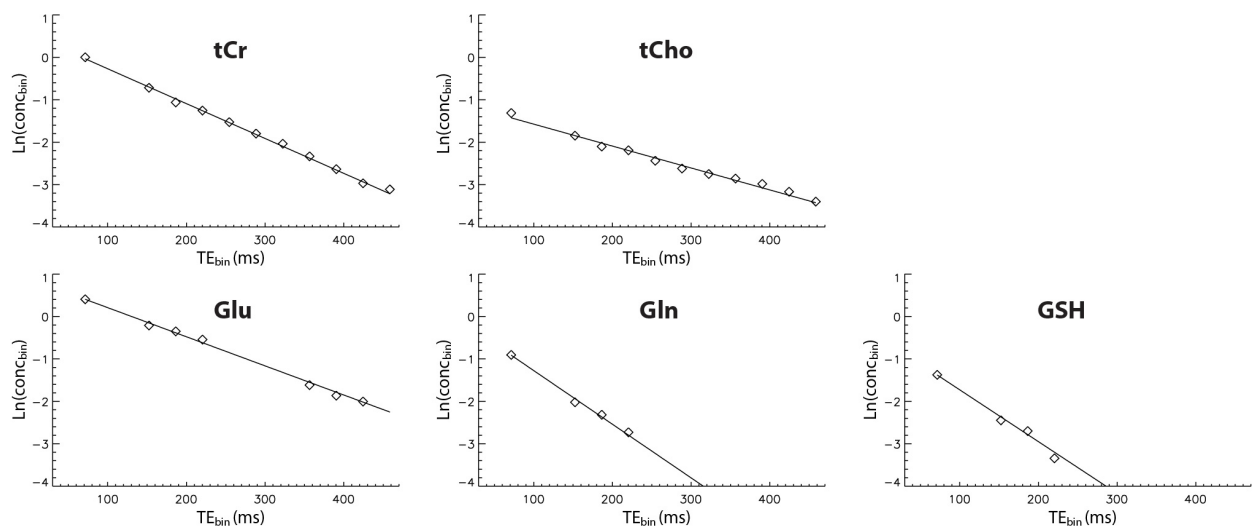

**Figure 6** Linear regression of ln(conc$_{bin}$) on TE$_{bin}$. The T$_2$-weighted concentrations (conc$_{bin}$) were obtained by fitting the bin spectra shown in Figure 4.

Linear regressions of ln(conc$_{bin}$) on TE$_{bin}$ for tCr, tCho, Glu, Gln, and GSH are shown in Figure 6. NAA was not analyzed as it was partially dephased by the 2.12 ppm band of the J-locking pulses and the crusher gradients. The T$_2$-weighted concentrations (conc$_{bin}$) were obtained by fitting the bin spectra shown in Figure 4. The single exponential model was found to fit the in vivo T$_2$ decay data well. The coefficients of determination ($R^2$) were greater than 0.95 for tCr, tCho, Glu, and Gln (see Table 1). Table 1 lists the neurochemical concentrations, as ratios to [tCr], and T$_2$ relaxation times obtained from the linear regressions.

|  | Neurochemical ratio (/[tCr]) | CRLB (%) | CV (%) | T$_2$ (ms) | CRLB (%) | CV (%) | $R^2$ |
|---|---|---|---|---|---|---|---|
| **tCr** | 1 | 1.1 ± 0.3 | 0 | 116 ± 6 | 0.6 ± 0.1 | 2.3 | 0.99 ± 0.00 |
| **tCho** | 0.21 ± 0.01 | 1.2 ± 0.3 | 2.4 | 182 ± 12 | 0.8 ± 0.2 | 2.7 | 0.99 ± 0.01 |
| **Glu** | 1.38 ± 0.12 | 2.1 ± 0.6 | 5.6 | 142 ± 11 | 1.8 ± 0.5 | 3.9 | 0.97 ± 0.03 |
| **Gln** | 0.52 ± 0.09 | 9.3 ± 2.4 | 7.1 | 77 ± 8 | 6.1 ± 1.9 | 4.0 | 0.95 ± 0.06 |
| **GSH** | 0.24 ± 0.07 | 20.8 ± 16.1 | 16.2 | 98 ± 17 | 15.1 ± 11.1 | 9.6 | 0.82 ± 0.19 |

**Table 1** Quantification of neurochemical concentrations (1/[tCr]) and T$_2$ relaxation times in the cingulate cortex of healthy participants (n = 6; mean ± SD). The voxel had 54.5% ± 5.1% grey matter, 37.9% ± 5.0% white matter, and 7.6% ± 4.4% cerebrospinal fluid. The within-subject coefficients of variation (CVs) were calculated from test and re-test measurements of the same voxel.

## Discussion

Cellular pathology in brain disorders has mainly been characterized by postmortem methods[12-15], which are difficult to standardize due to uncontrollable physicochemical and morphological changes in cells after death, complicating the understanding of disease etiology and development of effective treatment strategies. For example, although numerous postmortem studies have implicated glutamatergic dysregulation in central pathologies, progress in developing therapeutics targeting the glutamatergic system has been relatively slow. This is due, in part, to the lack of tools that can probe and monitor the pathophysiological processes in glutamatergic neurons and glia accompanying disease progression and treatment response.

MRS has high potential for generating cell type-specific contrast in vivo via molecule-environment interactions. However, the precision of measuring T$_2$ of J-coupled molecules using current MRS technologies has been inadequate. Furthermore, recent methodological re-examination of proton MRS, especially the commonly used short-TE MRS for neurochemical profiling, has found that spectral overlap between neurochemicals and the background signals, as well as between neurochemicals, can result in significant errors in both quantifying neurochemical concentrations and determining correlations between neurochemical concentrations and clinical metrics[16, 28, 29]. In particular, spectral overlap of the neurochemicals with the intense background signals in short-TE spectra was found to result in large errors in quantifying Glu, Gln, and GSH[16]. Although neurochemical concentrations measured by spectral fitting of short-TE MRS spectra have been treated implicitly or explicitly as spectrally uncorrelated variables in clinical MRS studies, severe spectral overlap can confound determination of correlations of biological origin[29]. The JL-CSE technique overcomes these challenges by effectively resolving Glu, Gln, and GSH at 7 Tesla and minimizing the background signals via TE averaging[25, 26] and by acquiring data at relatively long TEs.

Conventional in vivo MRS often uses relatively long TE to minimize the problematic background signals. However, this approach suffers from unknown $T_2$ weighting, as changes in $T_2$ are associated with many diseases[18, 19, 21]. Different $T_2$ weightings are routinely encountered in clinical MRS literature, leading to many controversies in clinical MRS findings[19]. The JL-CSE technique can efficiently determine the targeted neurochemical $T_2$s from its 2D dataset and provide neurochemical concentrations that are free of $T_2$ weighting.

When conventional MRS is used to measure molecular $T_2$s, the dynamic range for encoding the $T_2$s of J-coupled molecules can be very limited. This is because it is necessary to detect J-coupled molecules at fixed TEs for optimal sensitivity and spectral resolution due to J-evolution and spectral overlap[22, 23]. With conventional MRS, the Glu peak was detectable up to a TE of 374 ms[30], while the Gln peak was detectable up to a TE of 130 ms[23]. In contrast, JL-CSE enables detection of Glu up to a TE of 458 ms (bin 10), and Gln up to a TE of 220 ms (bin 3). Therefore, JL-CSE significantly broadens the dynamic range for measuring the $T_2$ relaxation times of the targeted J-coupled molecules. Table 1 shows that the $T_2$ values of Glu and Gln measured by JL-CSE are, to the best of our knowledge, of unprecedented precision as both Glu and Gln undergo $T_2$ decay while maintaining high-amplitude singlet spectral structure due to the action of the J-locking pulses. The very high precision of Glu and Gln $T_2$s makes JL-CSE a viable technique for characterizing cellular pathophysiology of glutamatergic neurons and glia in vivo.

Although the JL-CSE dataset was acquired in a fashion similar to 2D NMR, no coherence or polarization transfer occurs in the JL-CSE experiment by design. Therefore, 2D Fourier transform of the JL-CSE dataset is not meaningful as no cross peaks would be produced. Since the coherence transfer yield in a typical 2D NMR technique such as correlation spectroscopy (COSY) is very low, as evidenced by the low cross peak to parent (diagonal) peak ratio in COSY spectra, we deliberately avoided the conventional 2D approach. In JL-CSE, spectral resolution enhancement was achieved through the selective homonuclear decoupling effect of the J-locking pulses. An advantage of our approach is that it frees up the entire row ($t_2$) dimension for $T_2$ encoding without the need to detect the J-coupled spins at fixed TEs.

Since the current JL-CSE sequence is designed to measure the concentrations and $T_2$s of Glu, Gln, GSH, tCr, and tCho, a parsimonious fitting range of 1.8 – 3.4 ppm is sufficient for reliable quantification of these neurochemicals. The NAA singlet at 2.01 ppm is only 33 Hz away from the 2.12 ppm band of the J-locking pulses, causing partial dephasing of the NAA singlet. Therefore, NAA was not included in the analysis. When JL-CSE is parameterized to measure lactate or other neurochemicals, NAA is not affected.

In summary, we have developed a novel MRS technique that overcomes the inherent difficulties in measuring $T_2$ of J-coupled neurochemicals in vivo. It achieves J-locked chemical shift encoding in each column of the acquired 2D dataset. Chemical shift and $T_2$ are separately encoded in different dimensions. The columns of the 2D dataset are averaged into a small number of bins with the J-splitting of the targeted spins fully eliminated or suppressed. The background signals in the bin spectra are minimized by TE averaging and $T_2$ decay. The dynamic range for measuring the T$_2$s of the targeted spins is markedly broadened to 458 ms for Glu and 220 ms for Gln by the decoupling effect of the J-locking pulses. Furthermore, multi-band J-locking pulses can be used to decouple multiple molecules and suppress unwanted interfering signals.

As an application of JL-CSE, the concentrations and T$_2$ relaxation times of Glu, Gln, and GSH were measured. Bin spectra with well-resolved and sharp Glu, Gln, and GSH peaks were obtained with minimized spectral interference from the NAA aspartyl signals at 2.4 – 2.6 ppm. Glu and Gln T$_2$s were quantified with very high precision. JL-CSE therefore offers a viable noninvasive approach for establishing cell type-specific biomarkers of cellular pathophysiology in neuropsychiatric disorders in vivo.

## Methods

### Proof of Concept

A schematic diagram of a basic JL-CSE pulse sequence is shown in Figure 7a. This sequence is created by adding a 180° frequency-selective J-locking pulse at the midpoint between the two 180° refocusing pulses of a point-resolved spectroscopy (PRESS) sequence[31]. The J-locking pulse has a duration of 15 ms and its amplitude profile is generated by truncating a Gaussian function at one standard deviation on each side, leading to a full width half maximum (FWHM) bandwidth of 73 Hz. This J-locking pulse is applied at 4.10 ppm, targeting the H2 methine proton of lactate, to achieve J-locked chemical shift encoding for the H3 protons of lactate at 1.31 ppm. The timing parameters $\tau_1$, $\tau_2$, and $\tau_3$ are labeled in Figure 7a and their values are: $\tau_1$ = 17.5 ms, $\tau_3$ = 8.3 ms, and $\tau_2 = \tau_{2,0} + m\Delta\tau_2$, where $\tau_{2,0}$ = 17.5 ms, m is the $\tau_2$ increment number given by m = 0, 1, …, 255, and $\Delta\tau_2$ = 0.4 ms. The data acquisition window ADC$_1$ uses a dwell time of 0.2 ms and acquires 1060 data points. The pulse sequence uses 256 $\tau_2$ increments with fixed $\tau_1$ and $\tau_3$.

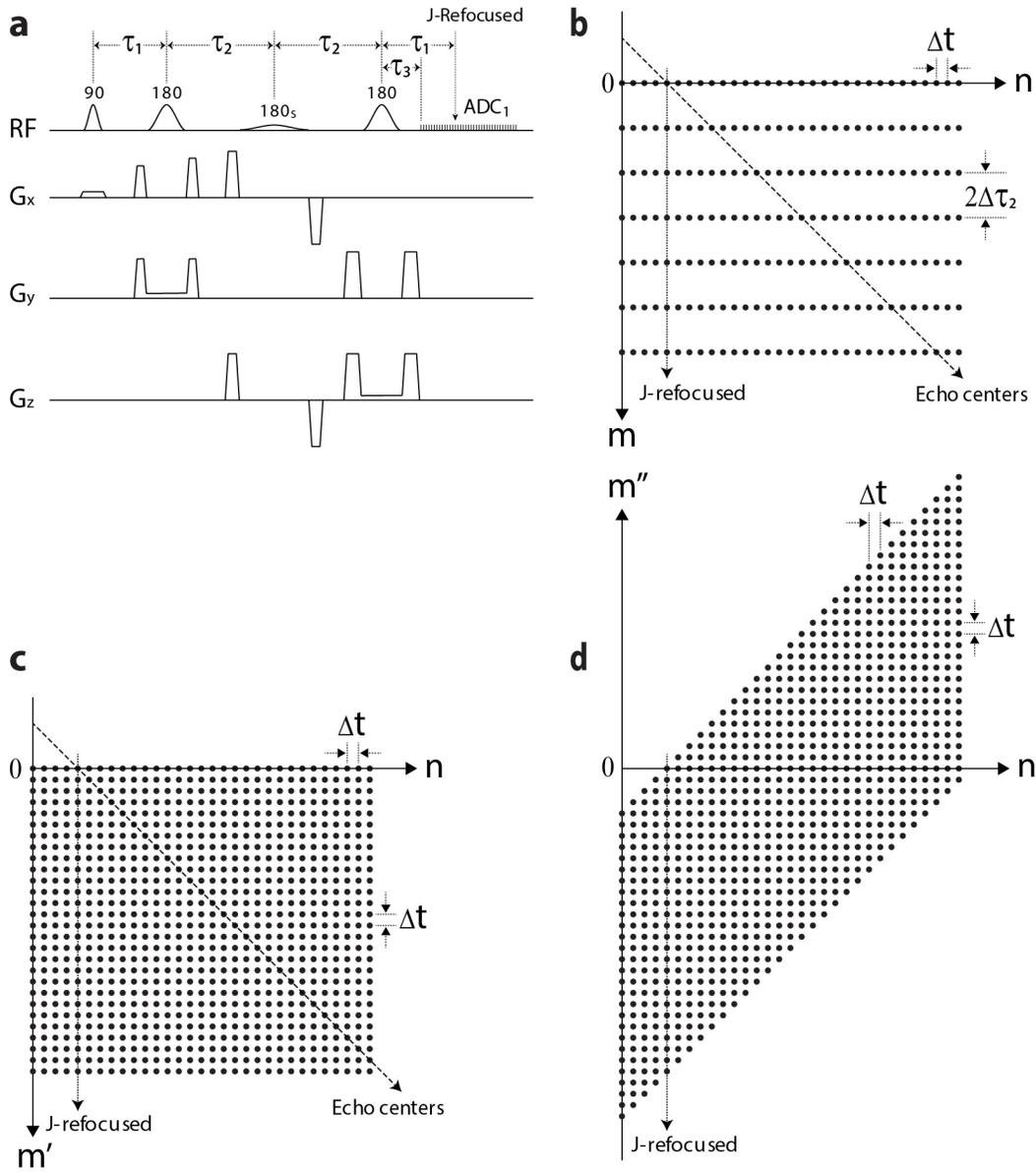

**Figure 7 a**. Schematic diagram of a basic JL-CSE pulse sequence. The frequency-selective J-locking pulse is denoted as $180_S$. $\tau_1 = 17.5$ ms; $\tau_2 = \tau_{2,0} + m\Delta\tau_2$, where $\tau_{2,0} = 17.5$ ms, m = 0, 1, … 255, and $\Delta\tau_2 = 0.4$ ms; $\tau_3 = 8.3$ ms; duration of $180_S = 15$ ms; dwell time $\Delta t = 0.2$ ms. **b**. The 2D dataset in the (m, n) space. **c**. The 2D dataset in the (m', n) space after sinc-interpolation. **d**. The 2D dataset in the (m'', n) space after aligning the echo centers.

A 256 × 1060 2D dataset was simulated for 256 $\tau_2$ increments using the pulse sequence shown in Fig. 7a. A schematic diagram of the timing of the data points is shown in Figure 7b. The row number m and column number n are labeled using zero-based numbering. The chemical shift encoding (CSE) time $t_{CSE}$(m, n) for data point (m, n), which is the time delay between the echo center at $4\tau_2$ and the $n^{th}$ data point at $\tau_1 + 2\tau_2 + \tau_3 + n\Delta t$ for the $m^{th}$ $\tau_2$ increment, is given by:

$$t_{CSE}(m, n) = \tau_1 + \tau_3 - 2\tau_{2,0} + n\Delta t - 2m\Delta\tau_2. \tag{1}$$

This shows that the column-wise time interval is $2\Delta\tau_2$ (0.8 ms), which is equal to $4\Delta t$.

For efficient data processing, the column-wise resolution is increased by a factor of 4 using sinc interpolation. After sinc interpolation, the dataset size in the (m', n) space becomes 1024 × 1060 and the time intervals in both dimensions are $\Delta t$, as shown in Figure 7c. The CSE time $t_{CSE}$(m', n) is the time delay between the echo center at $4\tau_2$ and the $n^{th}$ data point at $\tau_1 + 2\tau_2 + \tau_3 + n\Delta t$ in the $m'^{th}$ row of the sinc-interpolated dataset. It is given by:

$$t_{CSE}(m', n) = (n - m' - n_0)\Delta t, \tag{2}$$

where $n_0 = (2\tau_{2,0} - \tau_1 - \tau_3)/\Delta t = 46$. By setting $t_{CSE}$(m', n) to zero, the equation for echo centers is $n - m' - n_0 = 0$, forming the 45° diagonal line shown in Figure 7c. The row number for the echo center in the $n^{th}$ column is given by:

$$m'_{ec}(n) = n - n_0. \tag{3}$$

The TE value for the echo center in the $n^{th}$ column, which is the time delay between the excitation pulse and the echo center in row $m'_{ec}(n)$, is given by:

$$TE(n) = 2\tau_1 + 2\tau_3 + 2n\Delta t. \tag{4}$$

Each column in the (m', n) space is then shifted upwards by m'$_{ec}$(n) data points such that the echo centers of all columns lie on the horizonal n-axis. This creates a new dataset in the (m", n) space, where the new row number m" has both positive and negative values, with the positive direction being upward (see Figure 7d). Note that the positive m" direction in Figure 7d corresponds to the negative m' direction in Figure 7c. The void data points in Figure 7d are filled with zeros. In the (m", n) space, the CSE time for data point (m", n) becomes independent of the column number n and is given by:

$$t_{CSE}(m") = m"\Delta t. \tag{5}$$

This shows that the chemical shift information is Fourier-encoded in each column.

Due to the frequency-selective 180° pulse at $\tau_1 + \tau_2$ targeting spin S of the lactate I$_3$S spin system, the J-evolution of spin I is fully refocused at $2\tau_1 + 2\tau_2$, which is $\tau_1 - \tau_3$ after the start of data acquisition for all $\tau_2$ increments. The column number for the data points with fully refocused J-evolution is n$_{JR}$ = ($\tau_1 - \tau_3$) / $\Delta t$ = 46. Using the product operator formalism[32-34], the J-modulation factor of spin I at data point (m, n) is:

$$f_{JM}(n) = \cos[\pi J(n - n_{JR})\Delta t], \tag{6}$$

indicating a constant J-modulation factor in each column and therefore achieving J-locked chemical shift encoding in the column dimension.

To suppress the background signals and improve data processing efficiency, the columns of a dataset were averaged into a small number of bins. Each bin spectrum contains a frequency dependent Bloch-Siegert phase shift function[35] due to the frequency-selective J-locking pulse. The Bloch-Siegert phase shift function is the same for both row and column dimensions. As a result, the Bloch-Siegert phase shift function for the bin spectra was calculated by density matrix simulation at a single $\tau_2$ value.

## The Second J-locking Pulse

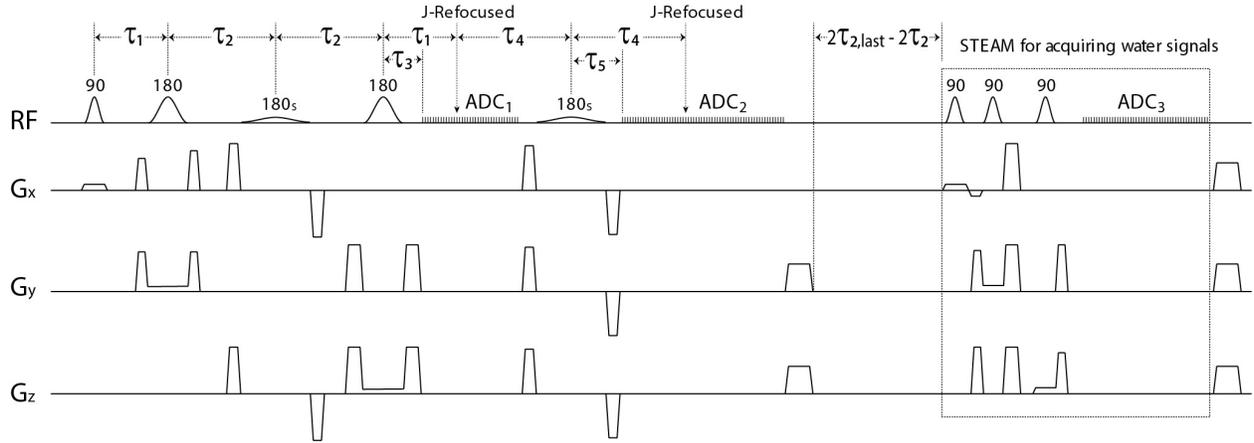

**Figure 8** Schematic diagram of the JL-CSE pulse sequence with a second J-locking pulse ($180_S$), a second ADC ($ADC_2$), and a STEAM block for acquiring unsuppressed water signals. The time delay between the main sequence block and the STEAM block is $2\tau_{2,last} - 2\tau_2$, where $\tau_{2,last}$ is the $\tau_2$ value for the last $\tau_2$ increment. $\tau_1 = 17.5$ ms; $\tau_2 = \tau_{2,0} + m\Delta\tau_2$, where $\tau_{2,0} = 17.5$ ms, $m = 0, 1, \ldots 255$, and $\Delta\tau_2 = 0.4$ ms; $\tau_3 = 8.3$ ms; $\tau_4 = 21.9$ ms; $\tau_5 = 10.9$ ms; duration of $180_S = 15$ ms; $ADC_1/ADC_2/ADC_3$ data points = 100/850/512; and $\Delta t = 0.2$ ms.

A second frequency-selective J-locking pulse is added to the basic pulse sequence in Figure 7a, as depicted in Figure 8. Implementation details of this pulse sequence are provided in the Supplementary Information. The number of data points for $ADC_1$ is reduced to 100 and the number of data points for $ADC_2$ is set to 850. The values for $\tau_1$, $\tau_2$, and $\tau_3$ values remain the same as in Figure 7a. The neurochemical signals acquired by $ADC_1$ and $ADC_2$ are combined and placed into a single 2D dataset in the (m, n) space. The time delay between the start of $ADC_1$ and the start of $ADC_2$ is $\tau_1 + \tau_4 + \tau_5 - \tau_3 = 110\Delta t$. Overall, the 2D dataset contains 256 rows and 1060 (100 + 110

+ 850) columns, in which columns 100 – 209 are filled with zeros. J-evolution for the targeted spins in column 265 (210 + ($\tau_4 - \tau_5$) / $\Delta t$) is refocused by the 2$^{nd}$ J-locking pulse.

A stimulated echo acquisition mode (STEAM)[36] block is appended to the main sequence block to acquire unsuppressed water signals for computing sensitivities of the multi-channel receiver coil[37] and frequency correction[38]. This STEAM block also pre-saturates the longitudinal magnetizations for the next $\tau_2$ increment.

**In Vivo Experiments**

Six healthy participants (4 female and 2 male; age = 44 ± 13 years) were recruited and scanned using a Siemens Magnetom 7 T scanner. Written informed consent was obtained from the participants before the study in accordance with procedures approved by our local institutional review board. A three-dimensional (3D) $T_1$-weighted magnetization prepared rapid gradient echo (MPRAGE) image was acquired with TR = 3 s, TE = 3.9 ms, data matrix = 256 × 256 × 256, and spatial resolution = 1 × 1 × 1 mm³. A 2 × 3.5 × 2 cm³ MRS voxel was placed in the cingulate cortex with a water linewidth of 13.4 ± 1.6 Hz. The JL-CSE pulse sequence (Figure 8) was used to acquire data with dual-band J-locking pulses. These pulses had a 180° band at 2.12 ppm and a 90° band at 4.38 ppm targeting the NAA aspartyl CH proton. Seven variable power RF pulses (sinc-Gauss pulse; duration = 26 ms; FWHM bandwidth = 105 Hz) with optimized relaxation delays (VAPOR) were used for water suppression.

The in vivo dataset in the (m", n) space was averaged into 11 bins with bin 0 containing all 100 columns of data from ADC$_1$ and the remaining 850 columns from ADC$_2$ evenly divided by bins 1 – 10. Details on $T_2$ quantification are provided in the Supplementary Information.

# References

1. Sibson, N.R. et al. In vivo 13C NMR measurements of cerebral glutamine synthesis as evidence for glutamate-glutamine cycling. *Proc Natl Acad Sci U S A* **94**, 2699-2704 (1997).
2. Shen, J. et al. Determination of the rate of the glutamate/glutamine cycle in the human brain by in vivo 13C NMR. *Proc Natl Acad Sci U S A* **96**, 8235-8240 (1999).
3. Yi, H., Talmon, G. & Wang, J. Glutamate in cancers: from metabolism to signaling. *J Biomed Res* **34**, 260-270 (2019).
4. Cluntun, A.A., Lukey, M.J., Cerione, R.A. & Locasale, J.W. Glutamine Metabolism in Cancer: Understanding the Heterogeneity. *Trends Cancer* **3**, 169-180 (2017).
5. Dringen, R., Brandmann, M., Hohnholt, M.C. & Blumrich, E.M. Glutathione-Dependent Detoxification Processes in Astrocytes. *Neurochemical Research* **40**, 2570-2582 (2015).
6. Ramadan, S., Lin, A. & Stanwell, P. Glutamate and glutamine: a review of in vivo MRS in the human brain. *Nmr in Biomedicine* **26**, 1630-1646 (2013).
7. Fazzari, J., Lin, H., Murphy, C., Ungard, R. & Singh, G. Inhibitors of glutamate release from breast cancer cells; new targets for cancer-induced bone-pain. *Sci Rep* **5**, 8380 (2015).
8. Das, T.K. et al. Antioxidant defense in schizophrenia and bipolar disorder: A meta-analysis of MRS studies of anterior cingulate glutathione. *Prog Neuro-Psychoph* **91**, 94-102 (2019).
9. Ekici, S., Risk, B.B., Neill, S.G., Shu, H.K. & Fleischer, C.C. Characterization of dysregulated glutamine metabolism in human glioma tissue with (1)H NMR. *Sci Rep* **10**, 20435 (2020).
10. Kennedy, L., Sandhu, J.K., Harper, M.E. & Cuperlovic-Culf, M. Role of Glutathione in Cancer: From Mechanisms to Therapies. *Biomolecules* **10** (2020).
11. Ekici, S. et al. Glutamine Imaging: A New Avenue for Glioma Management. *AJNR Am J Neuroradiol* **43**, 11-18 (2022).
12. Miladinovic, T., Nashed, M.G. & Singh, G. Overview of Glutamatergic Dysregulation in Central Pathologies. *Biomolecules* **5**, 3112-3141 (2015).
13. Rajkowska, G. & Miguel-Hidalgo, J.J. Gliogenesis and glial pathology in depression. *CNS Neurol Disord Drug Targets* **6**, 219-233 (2007).
14. Czeh, B. & Nagy, S.A. Clinical Findings Documenting Cellular and Molecular Abnormalities of Glia in Depressive Disorders. *Front Mol Neurosci* **11**, 56 (2018).
15. Kocahan, S. & Dogan, Z. Mechanisms of Alzheimer's Disease Pathogenesis and Prevention: The Brain, Neural Pathology, N-methyl-D-aspartate Receptors, Tau Protein and Other Risk Factors. *Clin Psychopharmacol Neurosci* **15**, 1-8 (2017).
16. Zhang, Y. & Shen, J. Effects of noise and linewidth on in vivo analysis of glutamate at 3 T. *J Magn Reson* **314**, 106732 (2020).
17. Frahm, J. et al. Localized proton NMR spectroscopy in different regions of the human brain in vivo. Relaxation times and concentrations of cerebral metabolites. *Magn Reson Med* **11**, 47-63 (1989).
18. Ongur, D. et al. T(2) Relaxation Time Abnormalities in Bipolar Disorder and Schizophrenia. *Magnetic Resonance in Medicine* **63**, 1-8 (2010).

19. Bracken, B.K., Rouse, E.D., Renshaw, P.F. & Olson, D.P. T-2 relaxation effects on apparent N-acetylaspartate concentration in proton magnetic resonance studies of schizophrenia. *Psychiatry Research-Neuroimaging* **213**, 142-153 (2013).
20. Kowalewski, J. & Maler, L. Nuclear spin relaxation in liquids: Theory, experiments, and applications, Edn. 2nd. (CRC Press, Boca Raton; 2019).
21. Kuan, E., Chen, X., Du, F. & Ongur, D. N-acetylaspartate concentration in psychotic disorders: T2-relaxation effects. *Schizophrenia Research* **232**, 42-44 (2021).
22. Choi, C.H. et al. Improvement of resolution for brain coupled metabolites by optimized H-1 MRS at 7 T. *Nmr in Biomedicine* **23**, 1044-1052 (2010).
23. An, L., Li, S. & Shen, J. Simultaneous determination of metabolite concentrations, T1 and T2 relaxation times. *Magn Reson Med* **78**, 2072-2081 (2017).
24. Landheer, K., Gajdosik, M., Treacy, M. & Juchem, C. Concentration and effective T(2) relaxation times of macromolecules at 3T. *Magn Reson Med* **84**, 2327-2337 (2020).
25. Hurd, R. et al. Measurement of brain glutamate using TE-averaged PRESS at 3T. *Magn Reson Med* **51**, 435-440 (2004).
26. Zhang, Y. & Shen, J. Simultaneous quantification of glutamate and glutamine by J-modulated spectroscopy at 3 Tesla. *Magn Reson Med* **76**, 725-732 (2016).
27. An, L. et al. Roles of Strong Scalar Couplings in Maximizing Glutamate, Glutamine and Glutathione Pseudo Singlets at 7 Tesla. *Frontiers in Physics* **10** (2022).
28. Hong, S., An, L. & Shen, J. Monte Carlo study of metabolite correlations originating from spectral overlap. *J Magn Reson* **341**, 107257 (2022).
29. Hong, S. & Shen, J. Neurochemical correlations in short echo time proton magnetic resonance spectroscopy. *NMR Biomed*, e4910 (2023).
30. Ganji, S.K. et al. T2 measurement of J-coupled metabolites in the human brain at 3T. *NMR Biomed* **25**, 523-529 (2012).
31. Bottomley, P.A. in Patent, Vol. US4480228A (General Electric Co, US 1984).
32. Sorensen, O.W., Eich, G.W., Levitt, M.H., Bodenhausen, G. & Ernst, R.R. Product Operator-Formalism for the Description of Nmr Pulse Experiments. *Progress in Nuclear Magnetic Resonance Spectroscopy* **16**, 163-192 (1983).
33. Vandeven, F.J.M. & Hilbers, C.W. A Simple Formalism for the Description of Multiple-Pulse Experiments - Application to a Weakly Coupled 2-Spin (I= 1/2) System. *Journal of Magnetic Resonance* **54**, 512-520 (1983).
34. Packer, K.J. & Wright, K.M. The Use of Single-Spin Operator Basis-Sets in the Nmr-Spectroscopy of Scalar-Coupled Spin Systems. *Molecular Physics* **50**, 797-813 (1983).
35. Emsley, L. & Bodenhausen, G. Phase-Shifts Induced by Transient Bloch-Siegert Effects in Nmr. *Chem Phys Lett* **168**, 297-303 (1990).
36. Frahm, J., Merboldt, K.D. & Hanicke, W. Localized Proton Spectroscopy Using Stimulated Echoes. *Journal of Magnetic Resonance* **72**, 502-508 (1987).
37. An, L., van der Veen, J.W., Li, S.Z., Thomasson, D.M. & Shen, J. Combination of multichannel single-voxel MRS signals using generalized least squares. *Journal of Magnetic Resonance Imaging* **37**, 1445-1450 (2013).
38. An, L., Araneta, M.F., Johnson, C. & Shen, J. Effects of carrier frequency mismatch on frequency-selective spectral editing. *Magnetic Resonance Materials in Physics Biology and Medicine* **32**, 237-246 (2019).

## Acknowledgements

We thank Christopher S. Johnson, MS, Maria Ferraris Araneta, CRNP, Tara Turon, CRNP, and Inna Loutaev, DNP & CRNP, for generous and valuable help. This study (NCT01266577) was supported by the Intramural Research Program of the National Institute of Mental Health, National Institutes of Health (IRP-NIMH-NIH, ZIAMH002803).

## Author Contributions

J.S. and L.A. conceived the technique and prepared the manuscript. L.A. developed the technology and performed the experiments.

# Supplementary Information

### In Vivo Re-Test Spectra and $T_2$ Quantification

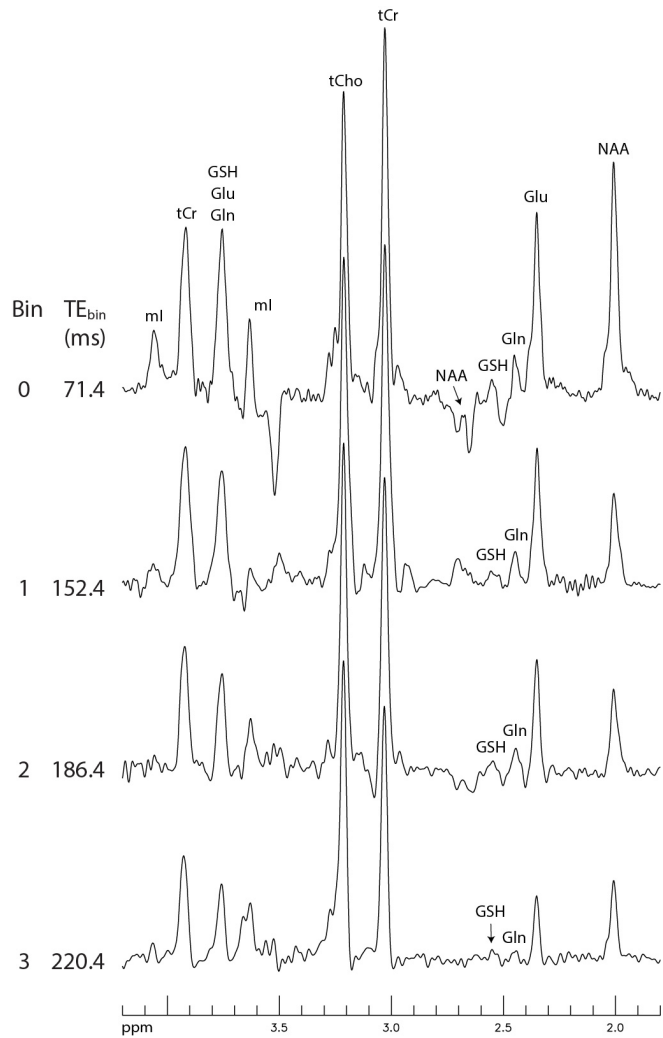

**Figure S1** In vivo spectra obtained from the re-test measurement of the spectra shown in Figure 3.

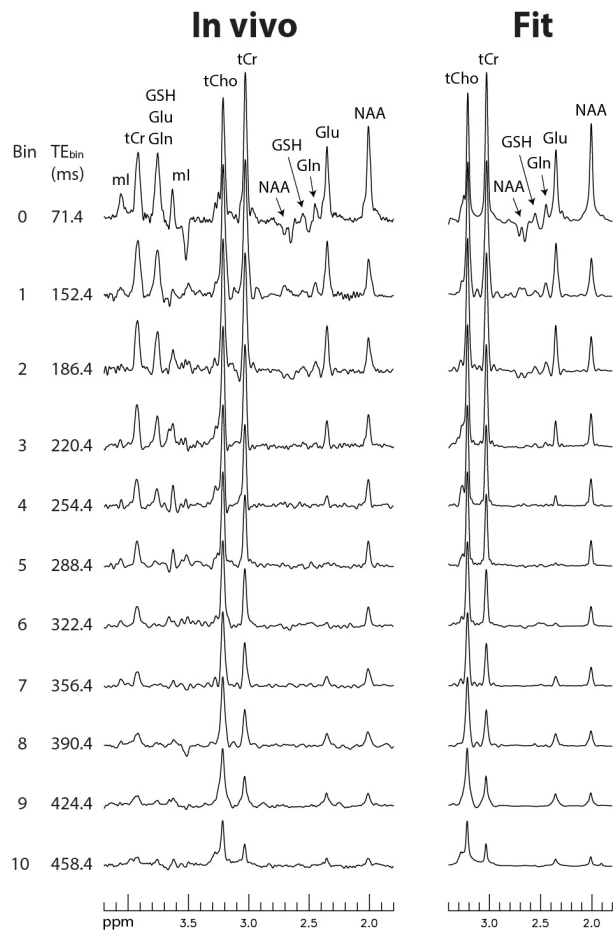

**Figure S2** Re-test of Figure 4 results.

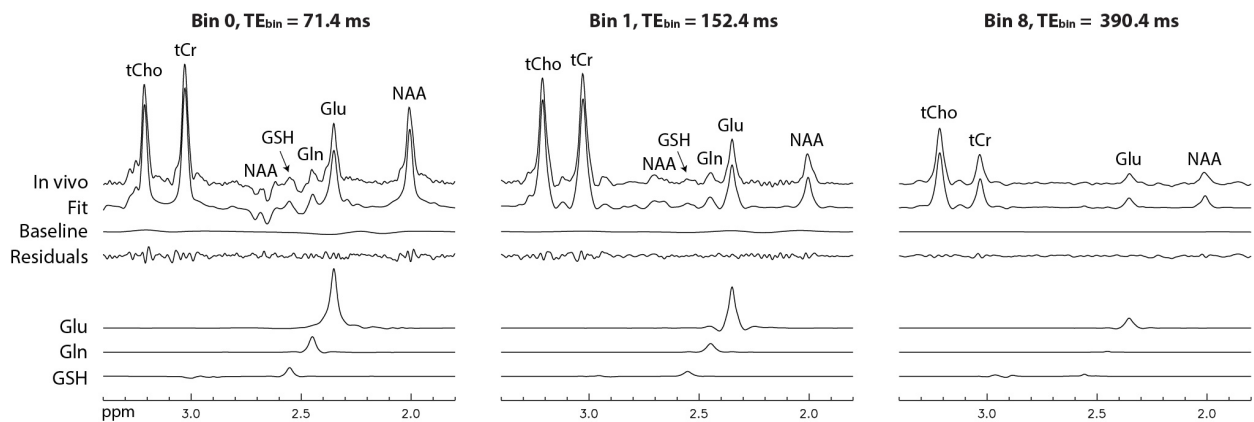

**Figure S3** Fitting details for spectra of 0, 1, and 8 shown in Figure S2.

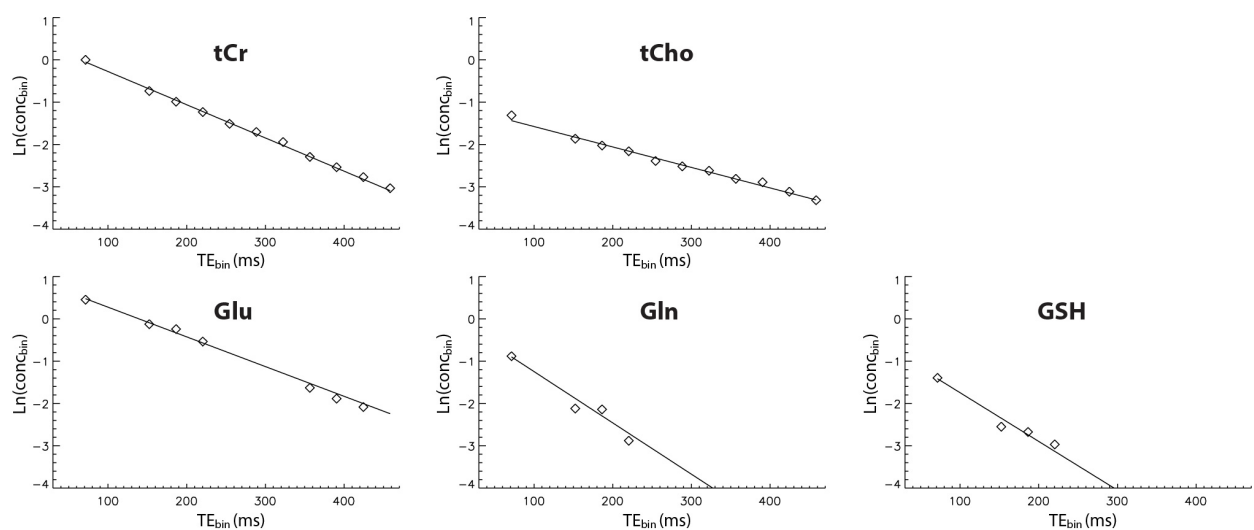

**Figure S4** Linear regression analysis of the spectral data shown in Figure S2.

**T$_2$ Relaxation**

The T$_2$ decay factor for data point (m", n) is $\exp[-(TE(n) - m"\Delta t) / T_2]$, which can be factorized into $\exp[-(TE(n) - TE_{bin}(h)) / T_2] \cdot \exp(m"\Delta t / T_2) \cdot \exp[-TE_{bin}(h) / T_2]$, where $TE_{bin}(h)$ is the average TE of all columns in bin h.

The first factor, $\exp[-(TE(n) - TE_{bin}(h)) / T_2]$, corrects for the small intra-bin T$_2$ decay effect. The second factor, $\exp(m"\Delta t/T_2)$, is independent of the column number and only results in a minor lineshape distortion because $\Delta t \ll T_2$. The third factor, $\exp[-TE_{bin}(h) / T_2]$, represents T$_2$ decay across different bins and therefore dominates the T$_2$ decay process. The T$_2$ and concentration of each molecule were obtained using weighted linear regression analysis[1] with an iterative fitting process. During the first iteration, the first two T$_2$ decay factors were ignored. Empirically, we found that two or three iterations were sufficient to achieve full convergence in T$_2$.

A two-sided Voigt function, $R \cdot \exp(-\lambda |m"|\Delta t) + (1 - R) \cdot \exp[-a(m"\Delta t)^2]$, was multiplied to the basis datasets in the (m", n) space to account for the line-broadening caused by the reversible

$T_2'$ relaxation effect[2], Here, R is the proportion of Lorentzian, $\lambda = \pi w$, $a = (\pi w/2)^2 / \ln(2)$, and w is the linewidth.

**Implementation Details of the JL-CSE Pulse Sequence Shown in Figure 8**

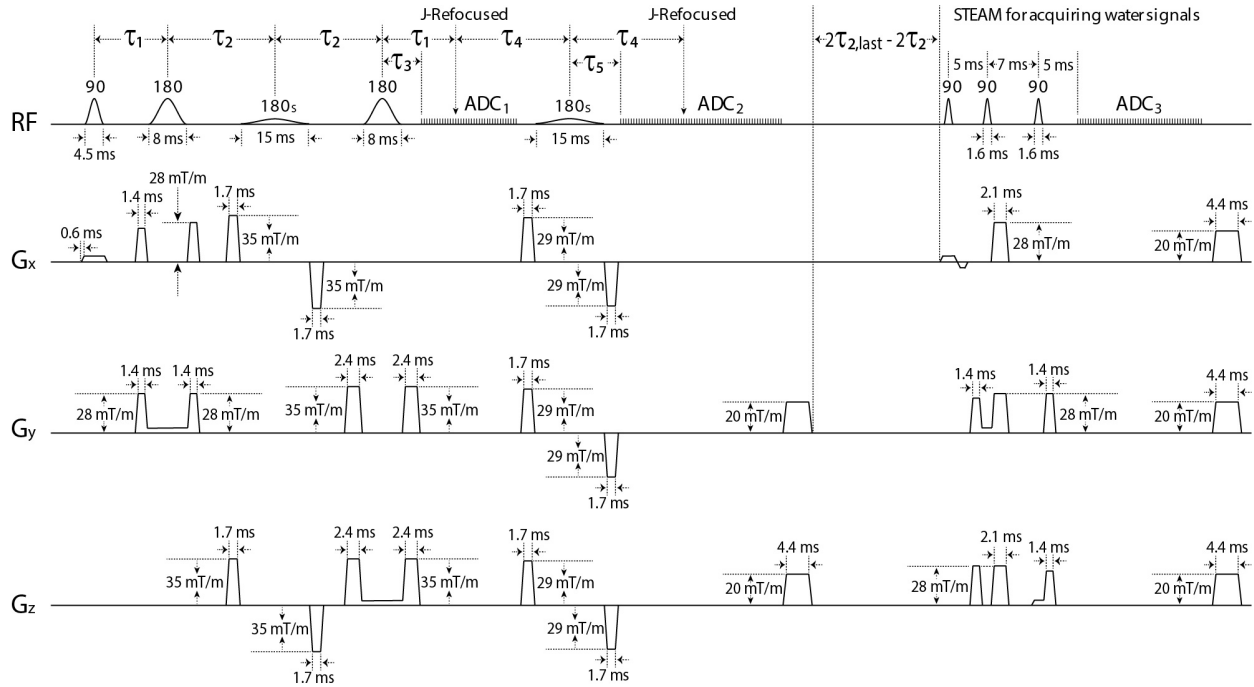

**Figure S5** Detailed timing diagram of the JL-CSE pulse sequence shown in Figure 8.

The slice-selective excitation pulse in the main sequence block was an asymmetric amplitude-modulated pulse[3] with a duration of 4.5 ms, $B_{1,max}$ of 18.6 µT, FWHM bandwidth of 3.1 kHz, and rephase fraction of 0.167. The slice-selective refocusing pulses were amplitude-modulated with a duration of 8.0 ms, $B_{1,max}$ of 18.6 µT, and FWHM bandwidth of 2.0 kHz. The J-locking pulses ($180_S$) had a duration of 15 ms and consisted of two bands: a 180° band at 2.12 ppm and a 90° band at 4.38 ppm. The three slice-selective excitation pulses in the STEAM block were a sinc-Gauss pulse with a duration of 1.6 ms, $B_{1,max}$ of 14.9 µT, and FWHM bandwidth of 2.8 kHz. The ramp-up and ramp-down times for all gradients were 0.6 ms. Other used parameters included: $\tau_1$ = 17.5 ms; $\tau_2 = \tau_{2,0} + m\Delta\tau_2$, where $\tau_{2,0}$ = 17.5 ms, m = 0, 1, … 255, and $\Delta\tau_2$ = 0.4 ms; $\tau_3$ = 8.3 ms; $\tau_4$ = 21.9 ms; $\tau_5$ = 10.9 ms; number of data points for $ADC_1/ADC_2/ADC_3$ = 100/850/512; $\Delta t$ = 0.2 ms; TR = 2.5 s; and total scan time = 10 min and 43 s.

**References**

1. Chatterjee, S. & Hadi, A.S. Regression Analysis by Example, Edn. 5th. (John Wiley & Sons).
2. Kowalewski, J. & Maler, L. Nuclear spin relaxation in liquids: Theory, experiments, and applications, Edn. 2nd. (CRC Press, Boca Raton; 2019).
3. Murdoch, J.B., Lent, A.H. & Kritzer, M.R. Computer-Optimized Narrow-Band Pulses for Multislice Imaging. *Journal of Magnetic Resonance* **74**, 226-263 (1987).